# Doped Biomolecules in miniaturized electric junctions

*Elad Mentovich [1,2], Bogdan Belgorodsky[2], Michael Gozin[2], Shachar Richter [1,3*] , Hagai Cohen,[4*]*

[1] Center for nanoscience and nanotechnology, School of Chemistry [2] Faculty of exact sciences[3], Tel-Aviv University, P.O. Box 39040, Tel-Aviv 69978, Israel.

[4] Department of Chemical Research Support, Weizmann Institute of Science, Rehovot, 76100, Israel .

Hagai.Cohen@weizmann.ac.il, srichter@post.tau.ac.il

**ABSTRACT** Control over molecular scale electrical properties within nano junctions is demonstrated, utilizing site-directed $C_{60}$ targeting into protein macromolecules as a doping means. The protein molecules, self-assembled in a miniaturized transistor device, yield robust and reproducible operation. Their device signal is dominated by an active center that inverts affinity upon guest incorporation and thus controls the properties of the entire macromolecule. We show how the leading routs of electron transport can be drawn, spatially and energetically, on the molecular level and, in particular, how the dopant effect is dictated by its 'strategic' binding site. Our findings propose the extension of microelectronic methodologies to the nanometer scale and further present a promising platform for ex-situ studies of biochemical processes.

**KEYWORDS** (Word Style "BG_Keywords"). Molecular electronics; Bio electronics; nano doping; Chemically

Extensive efforts are continuously devoted to the understanding and development of soft matter electronics [1-4] towards using organic molecules as flexible building blocks in dense architectures. [5, 6] The need in fine-control over their optical and electrical properties, e.g. via doping, [7, 8] presents a major challenge, since statistical dopant distribution is applicable only to components much larger than the average distance between neighboring dopants. [9, 10] Alternative approaches are therefore needed, e.g. by directing dopant species to pre-selected sites of the host matrix.

In this respect, macromolecules [11-14] and solid-state biomolecules in particular [15-17] offer useful advantages. Upon hosting foreign species, fine changes in the macromolecule properties can potentially be achieved with minimal effect on their structural and assembly characteristics. Nature often utilizes the binding of small molecules at host sites, e.g. in the transport of hydrophobic molecules through lipid-binding protein complexes, [18] and of fatty acids by albumins. [19] Here we exploit this feature for *electronic* applications, aiming at doping of nanoscale electric junctions and transistors. The electrical properties of the system are explored utilizing a nm-scale Central Gate Vertical Molecular Transistor [20-22] (C-Gate MolVeT, Fig. 1, bottom left) and a contactless technique, Chemically Resolved Electrical Measurements (CREM, Fig. 1, bottom, right) [23-26], which can resolve the local potential at selected chemical addresses. Intriguing details of the electrical transport across the macromolecule and, specifically, the role of the $C_{60}$ guest in switching the molecule electrical affinity are thus revealed.

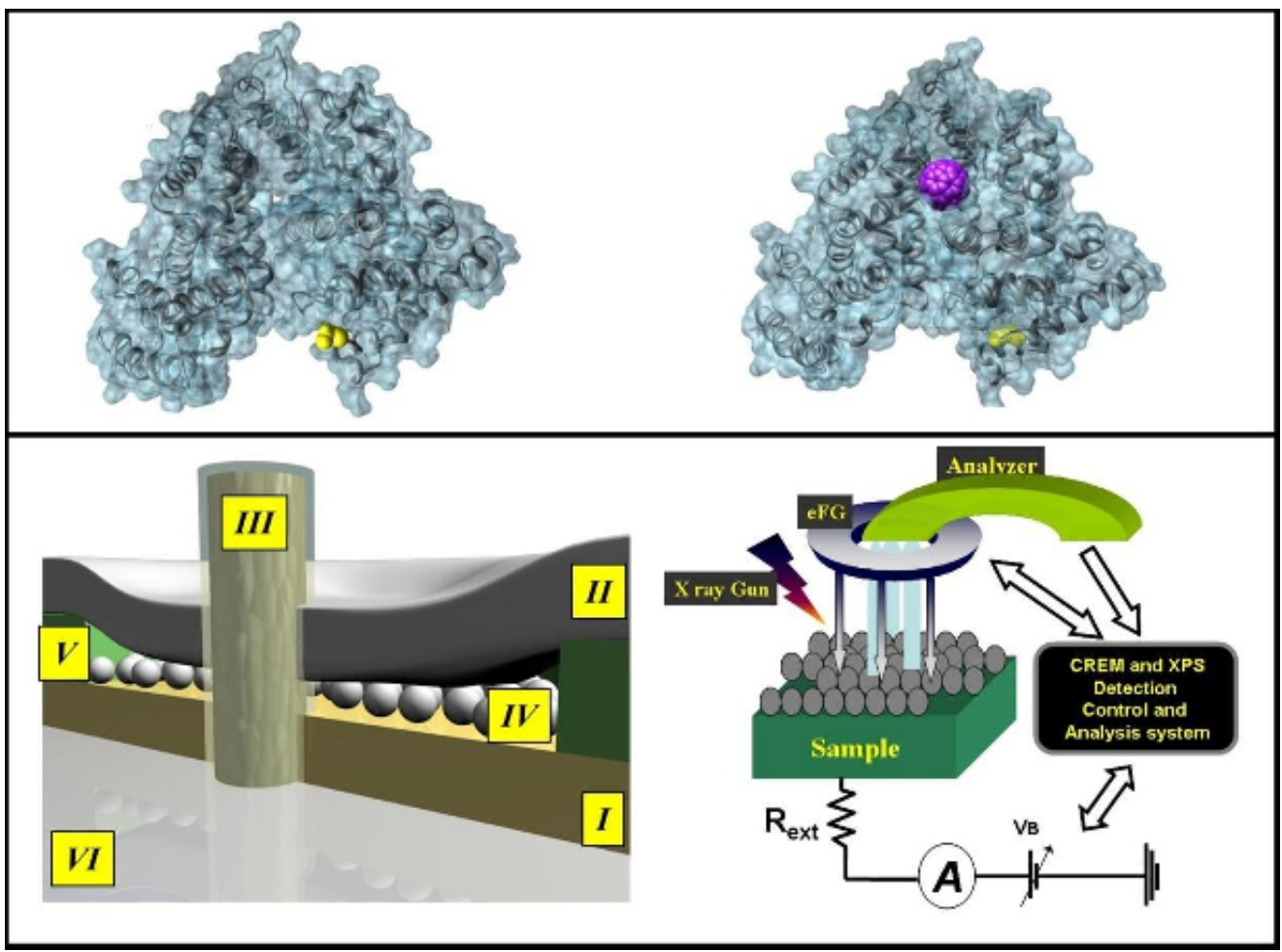

**Figure 1**: Top: Schematic illustration of the un-doped (left) and $C_{60}$-doped (right) BSA. Bottom left: A simplified cartoon of the C-Gate MolVet structure. The vertical transistor is formed inside a microcavity: source (Au, I) drain (Pd, II) and gate (Ti-TiO2, III) electrodes that are used to activate the protein monolayer (IV). The microcavity is bordered by $Si_3N_4$ (V) and the gate electrode is activated by highly-doped silicon/Silicon oxide layer (VI). Bottom right: The CREM setup. Input signals are the x-ray irradiation, the eFG low-energy electrons and the sample bias. Output signals are the photoelectron spectrum and the sample current (see S.I. for details).

A simple synthetic methodology was used to get site-specific targeting of $C_{60}$ molecules into self-assembled monolayers (SAM) of Bovine Serum Albumin (BSA) protein (Figure 1, top).[27] BSA-$C_{60}$ is a well-defined complex, with approximately the same size as the original BSA protein. The $C_{60}$ bucky-ball has already been shown to bind selectively to an albumin site at subdomain IIA,[27] close to the Trp214 Tryptophan site.[28] Self-assembly on gold resulted in rather uniform BSA monolayers, ~4 nm thick, with a slightly more 'open' conformation (less compact organization) of the doped-BSA (see S.I. for synthesis, purification and characterization details).

Direct transport measurements were performed in a C-Gate MolVet device [20](Figure 1, bottom left), recording the current/voltage characteristics through the molecular layer, between source and drain electrodes, $I_{SD}/V_{SD}$, while modulating the field via a third, central gate electrode ($V_G$). This type of devices offers very high sensitivity at relatively low $V_{SD}$ and $V_G$ regimes, thus allowing efficient and non-destructive scanning of sample's molecular orbitals. [22, 29] Figure 2 top compares representative $I_{SD}/V_{SD}$ characteristics of the doped and undoped BSA-based junctions, measured at 77K. In both cases, the gate effect under positive $V_{SD}$ is negligible, in contrast to the negative $V_{SD}$ regime. At negative polarity of the $C_{60}$-BSA transistor current values are considerably lower than those of the undoped device, while similar current magnitudes are obtained for positive $V_{SD}$.

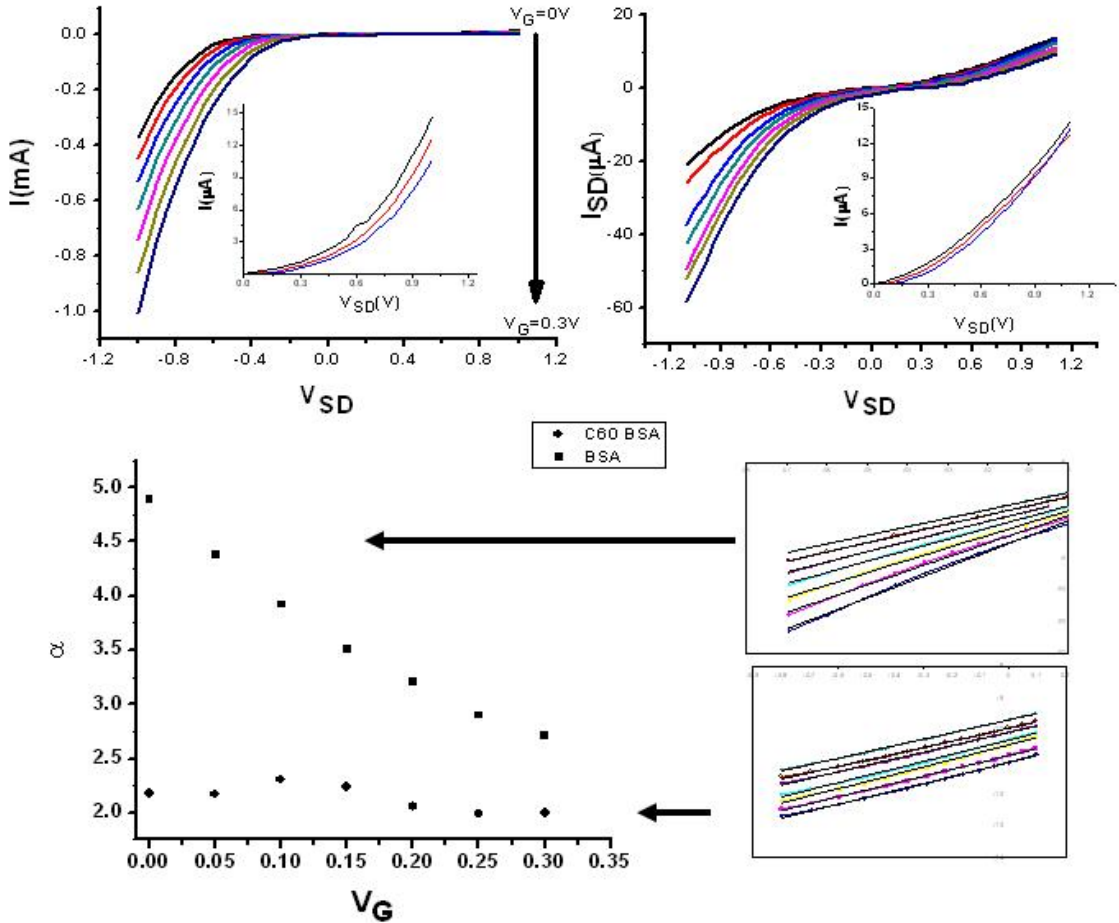

**Figure 2**: Transistor transport data. Top. $I_{SD}/V_{SD}$ of the un-doped (left) and the doped-BSA (right) device, as a function of $V_G$. Main panels (insets) show I-V curves recorded under negative (positive) $V_{SD}$. Note how the current magnitude increases (decreases) under negative (positive) polarity as $V_G$ is increased, pointing at electron (hole) dominated transport mechanisms. Bottom. Variation of the negative-polarity exponent factor (α) with $V_G$ (left) in BSA (squares) and $C_{60}$-BSA (diamonds), as derived from the log-log I-V plots on the right.

As already discussed previously [20], Fowler-Nordheim tunnelling (FN) dominates the low-field negative $V_{SD}$ region. At the high-field regime, the curves obey $J \propto F^\alpha$, pointing at charge limited (CL) mechanisms [30], where $\alpha$ is the CL exponent. The critical field, at which CL begins to dominate over FN, is higher by ~250 mV in the $C_{60}$-BSA complex, as compared to the un-doped protein (see also S.I). Fig. 2 bottom further shows that $\alpha$ is highly gate-dependent in the un-doped BSA, suggesting a significant energy distribution of charge trap levels within the layer. [31] On the other hand, the doped junction exhibits identical slopes, $\alpha \approx 2$, which points to the leading role of a dominant trap energy (fig 2 bottom) [30, 31], similar to reports on conjugated-polymer devices[31]. From the $I_{SD}/V_G$ dependence one learns that under negative polarity the conduction is *electron* dominated (current increasing with $V_G$), while for positive polarity it is dominated by *hole* transport (current decreasing upon $V_G$ increase).

A complementing view on the studied systems is provided by the element-specific Chemically Resolved Electrical Measurement (CREM) curves, Fig. 3a-c, recorded from monolayers on gold substrates with no top contacts. The experiment utilizes photoelectrons to read electrostatic potentials from selected atomic sites. [23] It is conducted at room temperature, exposing the sample to a flux of slow (<3 eV) electrons under fixed source (eFG, see SI) conditions, and varying the bias on the sample ($V_B$) in a step-wise manner. For each step, both the sample current (I) and the potential changes (ΔV, as derived from shifts in the photoelectron kinetic energies, relative to measurement under minimal charging conditions) are recorded. [23, 26]

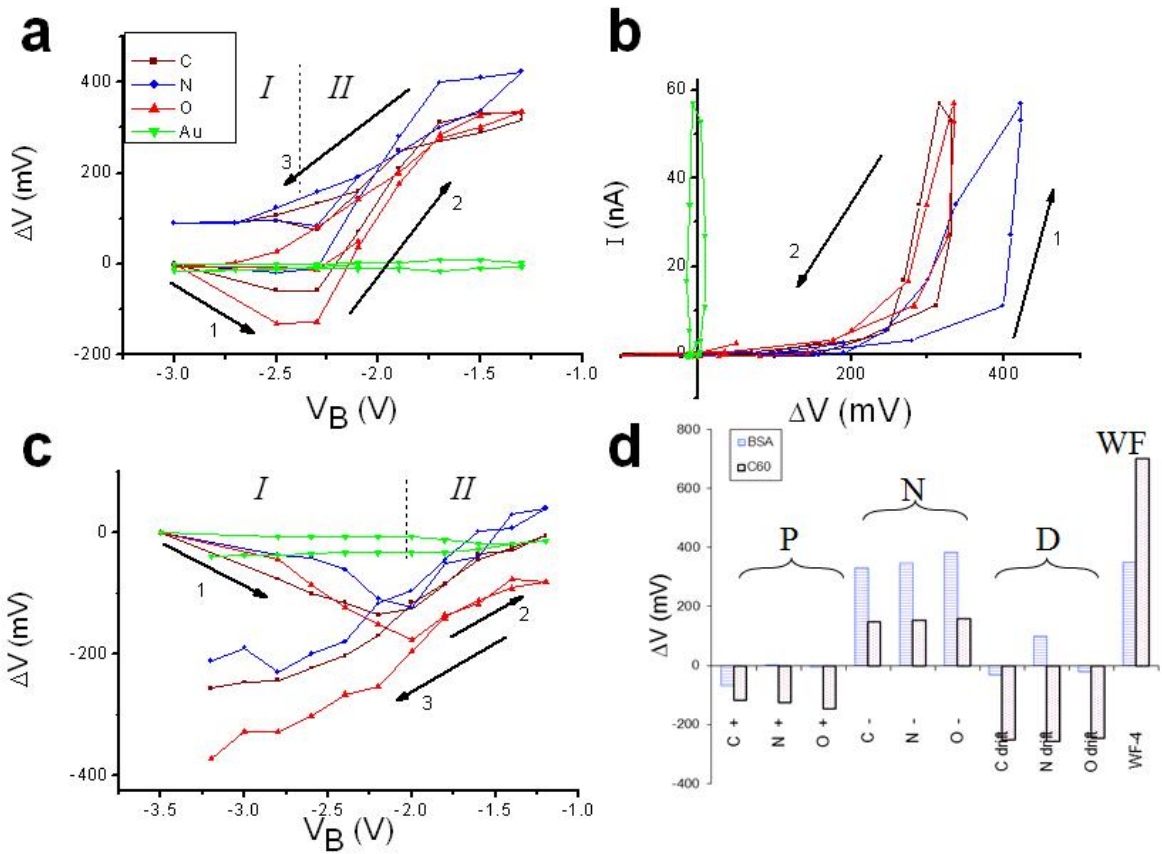

**Figure 3**: Element specific CREM data: (a) Surface potential (ΔV) *vs* sample bias (V$_B$) characteristics of undoped 40Å thick BSA monolayer on gold. Scan direction is given by the arrows. Two regimes are distinguished: (*I*) positive charging and (*II*) negative charging; (b) the corresponding current-voltage I-ΔV curves; (c) the C$_{60}$-BSA ΔV-V$_B$ curves; (d) a summary of: potential variations under positive (P) and negative (N) charging conditions; the electrostatic irreversibility (D, drift) along the first scan; the work-function of the fresh surface, shifted by -4 eV (WF). Note the reduced resistance and increased hysteresis in the doped system. The experimental error in ΔV is: 5 mV, 40 mV, 50 mV and 70 mV for the Au, C, O and N, respectively.

Two regimes are indicated in figures 3a,c: (*I*) A highly negative V$_B$ range, where the eFG electrons are totally repelled by the biased sample and the layer tends to accumulate x-ray induced positive charge, Q>0 [32]. (*II*) A low negative bias range, where eFG electrons are injected to the sample and negative surface charge (Q<0) is accumulated. The crossover between the two regimes, given by an inflection point (IP), was shown to define the work-function (WF) of the sample. [33]

Notably, significant ΔV values develop at the bare BSA, tending to saturate at high currents (low negative V$_B$) and undergoing hysteresis under decreasing the current back to low values. Differences between the elemental curves in Figs. 3a,b reflect affinity variations [34] and, in particular, enhanced mean affinity of the N sites to injected electrons and of the O sites to positive charge, to be discussed elsewhere. Note that the Au line-shifts determine the full back-contact impedance to be eliminated from the overlayer data.

The C$_{60}$-BSA curves in Fig. 3c manifest a marked doping effect: (1) enhanced positive charging; (2) reduced negative charging; and remarkably (3) asymmetric hysteresis: large irreversibility in all elemental curves at region *I*, while in region *II* all curves are fully reversible (no hysteresis). Thus, the C$_{60}$ molecules, known to be very good electron acceptors [35], do not tend to capture the injected electrons (no hysteresis in *II*). On the contrary, *hole* trapping becomes of very long life-time characteristics (pronounced hysteresis in *I*). Average potential variations and the overall irreversibility (electrostatic drift) are summarized in Fig. 3d.

To understand this striking result, note first that a ~350 meV difference in work function (WF) is extracted from the inflection points (between region *I* and *II*) in Figs. 3a,c, yielding WF=4.35 eV and 4.7 eV for the bare and the doped BSA, respectively. This WF change is associated with extended charge redistribution upon C$_{60}$ complexation (see discussion below), where the C$_{60}$ site becomes electron rich and a good donor, in agreement with the pronounced positive charging in Fig. 3c. Second, both WF values are considerably smaller than those of Au and Pd (~5.0 eV); hence, interface dipoles at the electrode contacts must be considered, acting both *against* electron conduction. In spite of these dipoles, dominant electron conduction is measured under negative polarity, which highlights the intrinsic medium property: highly favored electron transport. The interface dipoles necessarily differ in width (see Fig. 4a), a fact inferred from the rectification in Fig. 2 (see also SI for bare BSA data), such that under positive polarity, the thicker barrier at the Pd side leads to hole-dominated transport.

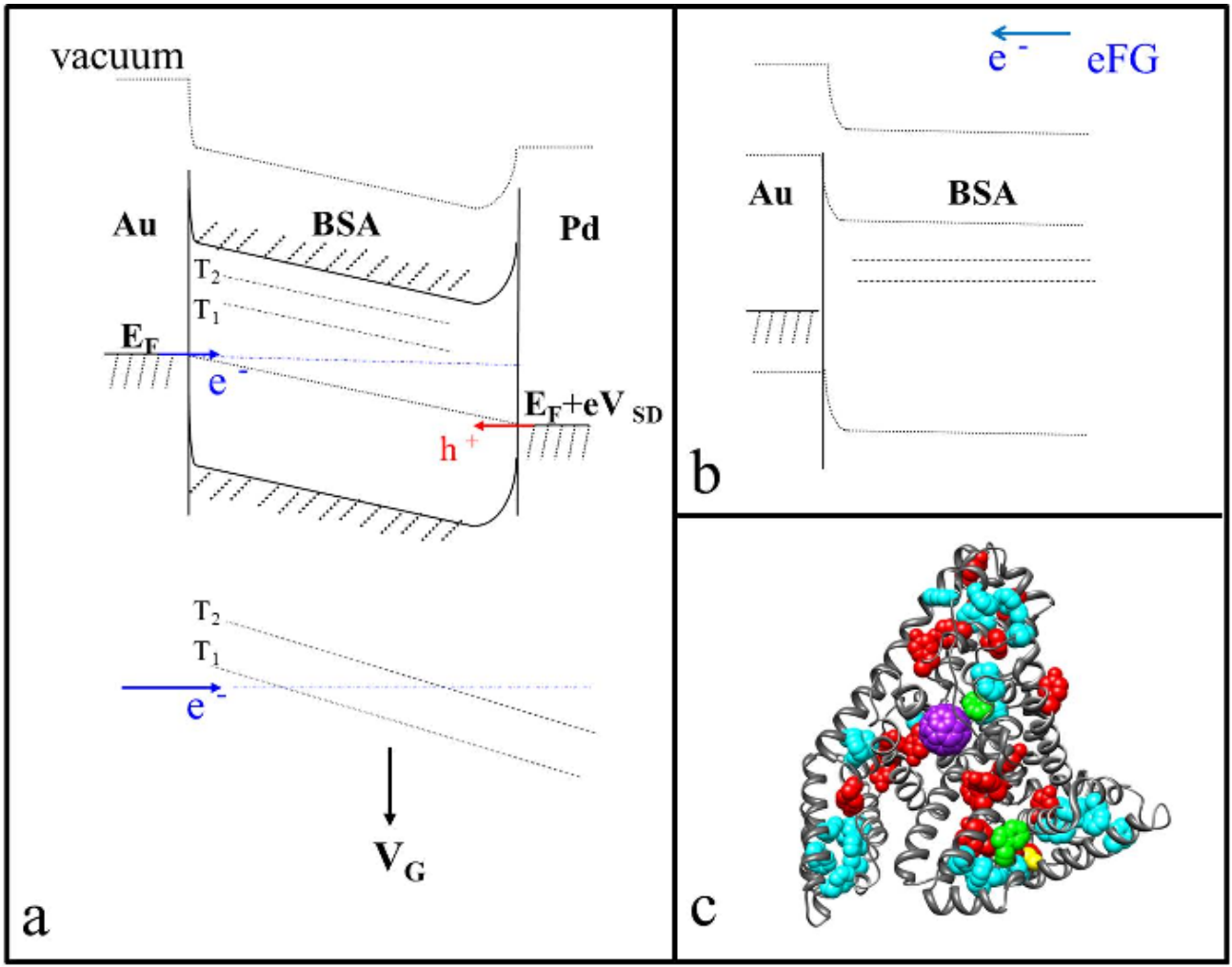

**Figure 4**: The transport mechanism: (a) A schematic description of energy levels under negative bias conditions: hole injection from the top Pd contact and electron injection from the gold substrate are indicated. Acceptor trap states in the range $T_1$-$T_2$ dominate the charge transport; their position relative to the electrodes' $E_F$ is affected by $V_G$, as illustrated at the bottom scheme. Note that the $C_{60}$ dopant blocks $T_1$ levels. (b) A corresponding scheme under the eFG electron injection in CREM. (c) The BSA molecular structure with its leading acceptor (Phenylalanine, light blue) and donor (Tyrosine, red) sites. The $C_{60}$ (violet), the Tryptophan sites (green) and the cysteine binding group (yellow) are also indicated. A 3D movie presentation of the molecule is provided in the SI section.

The CL transport dynamics in bare BSA likely involves plural trap energies, leading to high exponent values $\alpha \gg 2$ [36]. Fig. 2 shows however that α decreases rapidly (→2) for increased $V_G$, indicating that the corresponding energy width is limited, on the scale of 350 meV, as illustrated by $T_1$ and $T_2$ in Fig.

4a. Our results further suggest that upon complexation with $C_{60}$, critical $T_1$ sites transform from acceptors to strong donors and, thus, block electron conduction. This process is expressed (see Fig. 2) in (1) a drastic current suppression; (2) a strict $\alpha \approx 2$ exponent value; and (3) a shift of the crossover voltage between the FN and CL regimes. Accordingly, the doping effect becomes negligible under positive polarity, when *hole* transport dominates, which fully agrees with our argument on selective inactivation of acceptor-type hopping routes.

The above results can be better understood in light of the 3D molecular structure (see movie, S.I.). Fig. 4c shows a 2D projection of the BSA molecule with its leading acceptor (phenyl, light blue) and donor (tyrosine, red) groups. [37] Recalling that the empty (no $C_{60}$) subdomain IIA presents a strong acceptor site (tryptophan, Trp214) and that the CL conduction of electrons is carried by acceptors, an injected electron can easily complete a continuous hopping path (upward, through phenyl groups, blue) in the undoped case. However, as soon as the central Trp site turns to be a donor, the favored electron-hopping paths are blocked. Notably, it is not just the $C_{60}$ character but also its critical 'bottle-neck' location that dictates the electrical behavior and, indeed, the *hole* conductance (under positive polarity) is negligibly affected by the dopant, as expected from the proposed mechanism.

Identification of the leading conduction routes in the BSA macromolecule is an important outcome of this work. Fig. 4c further reveals another interesting feature: a broad spatial distribution of donor groups that collectively take part in charge donation to the $C_{60}$ site during complexation: contribution from distant tyrosine groups, including groups located close to the Au substrate, is essential for explaining the experimentally observed dipole sign and magnitude. In fact, the magnitude of the molecular dipole should be self-limited by the presence of acceptor states that can gain charge as soon as the intra-molecular potential-drop exceeds $T_2$-$T_1$ (Fig. 4a), which is in good agreement with our measurements (~350 meV). This very charge transfer further explains the inefficient discharge of holes from the $C_{60}$ site, manifested by Fig. 3c: First, it creates a molecular dipole that acts against hole discharge to the gold and, second, the leading candidate sites for hole-hopping are partially blocked ('emptied' from electrons).

Our results imply on the potential suitability of biological molecules to future electronic devices, exhibiting both stability and useful functionality. The inspected BSA layers are relatively very robust, far better than e.g. alkane chain monolayers,[10, 25] retaining reproducible and well-behaving appearance in the transistor device (up to a year already for $V_{SD}$ < 2 V; see SI for more details) and, to a lesser extent, under long x-ray irradiation. This may originate in their dry condensed phase, as opposed to aqueous environment in biological systems, where degradation is accelerated [11, 12, 38]. The recognition capabilities of biological molecules present another useful feature: BSA can absorb foreign species with minor influence on its assembly characteristics and yet with a marked impact on electrical properties. The present usage of recognition is very different from e.g. the self-wiring applications proposed for DNA. [39, 40] It is aimed at novel doping-like variability in molecule properties, playing with the host-guest combination and their associated binding site.

The observed switch in molecule affinity is essentially the function of an active center; here responsible for the behavior of a *macro* molecule. Can one learn from these experiments on biological mechanisms involving binding and release of small molecules? [41, 42] The conditions here do not quite imitate the aqueous biological environment and, obviously, the $C_{60}$ molecule is not released here from its host. Yet, one does succeed in these experiments to controllably deviate from charge neutrality and follow the site-selective charge transfer mechanisms. Thus, the present methodology proposes a unique view on molecular level chemical activity, which may be proven useful in studies of biochemical mechanisms.

In summary, we have demonstrated a new approach to doping-like electrical control suited for the molecular scale and inspected its function by complementing nanoscale-sensitive electrical probes. Using biological molecules embedded in a solid-state transistor, we exploited the $C_{60}$ recognition at a specific protein domain for achieving accurate, site-directed modifier of the monolayer dielectric properties. We have shown how the manipulated protein site can switch between an acceptor and a donor state, obeying external stimuli for charge neutrality violation, and how detailed understanding of the leading conduction paths can be gained at the submolecular level. The present approach can be exploited for the development of improved sensors and nano-scale devices and, possibly, for studies of complex electron transfer mechanisms in biological systems.

**Materials and Methods.**

**C-Gate MolVet fabrication**. A network of gold electrodes was defined on top of a highly doped silicon wafer covered with 100 nm thick thermal oxide, followed by the deposition of a 70 nm layer of $Si_3N_4$ dielectric material. Next, Arrays of microcavities, ranging from 800 nm to 1.5 μm in diameter were created by drilling holes via reactive ion etcher (RIE) through the entire layer down to the highly doped silicon substrate, followed by mild wet etching of several nanometers of the gold electrode. This undercut in the electrode provided space for oxide growth. This step was followed by the evaporation of a titanium column, a photolithography shape definition of the larger cavity, and oxidation of the titanium column that formed the gate electrode. Self-assembly of the protein-based monolayer on top of the exposed gold ring was then performed, followed by shape definition of the upper electrode and indirect or chopper evaporation of palladium on top of the protein layer. Measurements were performed using cryogenic probe station equipped by semiconductor parameter analyzer (Keithley 4200 SCS). See SI for repeatability and performance characterization.

**Materials**: Essential fatty acids and globulin free BSA, fine chemicals, and solvents were purchased from Sigma–Aldrich. $C_{60}$-fullerene was purchased from Sesres, and γ-CD2C60 was synthesized according to previously reported methods. [43]

**BSA–$C_{60}$ complex preparation**: 50 μM solution of BSA in tris-acetate buffer (20 μM, pH 7.2) was incubated with two equivalents of γ-CD2C60 at 10 °C for 24 h. The complexation solution was separated and purified on a Sephadex G-25 gel-permeation column (Pharmacia Biotech) with tris-acetate buffer (20 mM, pH 7.2) or on HPLC TSK-GEL column. An optimized stepwise, removal-addition procedure included incubating BSA and γ-CD2C60 at 10 °C for 48 h at a 1:2 molar ratio in tris-acetate buffer (20 mM, pH 7.2), followed by the removal of γ-CD-fullerene clusters and γ-CD by size-exclusion chromatography (Pharmacia, G25 cartridge, 20 mM tris-acetate buffer, pH 7.2), and subsequent addition of 2 equivalents of γ-CD2C60 to a BSA containing fraction. This procedure was repeated six times, once every 48 h. The buffer concentration was reduced by reloading the complex-containing fraction on a second Sephadex G-25 column and eluting the complex with a tris-acetate buffer (1.0 mM, pH 7.2). The resulting complex solution was lyophilized for storage and further experiments. After reconstitution in water, the complex concentration in solution was determined by UV-visible spectroscopy and a BioRad protein assay (Bio-Rad Lab).

**CREM (Chemically resolved electrical measurements):**

The XPS-based electrical measurements were performed on a slightly modified Kratos AXIS-HS setup, using monochromatic X-ray source, Al kα, at low power, 75 W, and base pressure of $1 \cdot 10^{-9}$ torr. The eFG was operated at 1.8 A filament current and -2.5 V grid bias. A Keithley 487 electrometer was connected to the sample back contact, providing both current detection and sample biasing. For reliable extraction of the layer dielectric response, reversible line-shifts were differentiated from any irreversible electrostatic (and chemical) modifications. Detailed follow-up of the beam induced changes was further conducted (see SI) such as to extract the degradation information on a broad range of time scales. Rapid CREM-based WF measurements were performed initially, before exposure to any irradiation, and later on again, both on fresh and on irradiated spots at sequencing stages of the experiment. (Early electrostatic changes could frequently be identified, but soon the electrical data stabilized and, also, no significant changes in the standard (XPS) chemical analysis were observed, indicating high stability of the system.) The error in ΔV determination can approach ≤5 mV; but for the noisy signals we achieved 30-70 mV.

ACKNOWLEDGEMENT: This research was supported by the Israeli science foundation, project #604, and the USAF.

**Supporting Information Available**. The supporting information incudes: (1) A note regarding doping at the nm scale; (2) Complementing aspects of the present electrical probes; (3) Technical comments on the CREM study; (4) Layer characterization; (5) Protein structure calculation method (movie included); (6) Information on the device performance.

**References and footnotes**

SYNOPSIS TOC

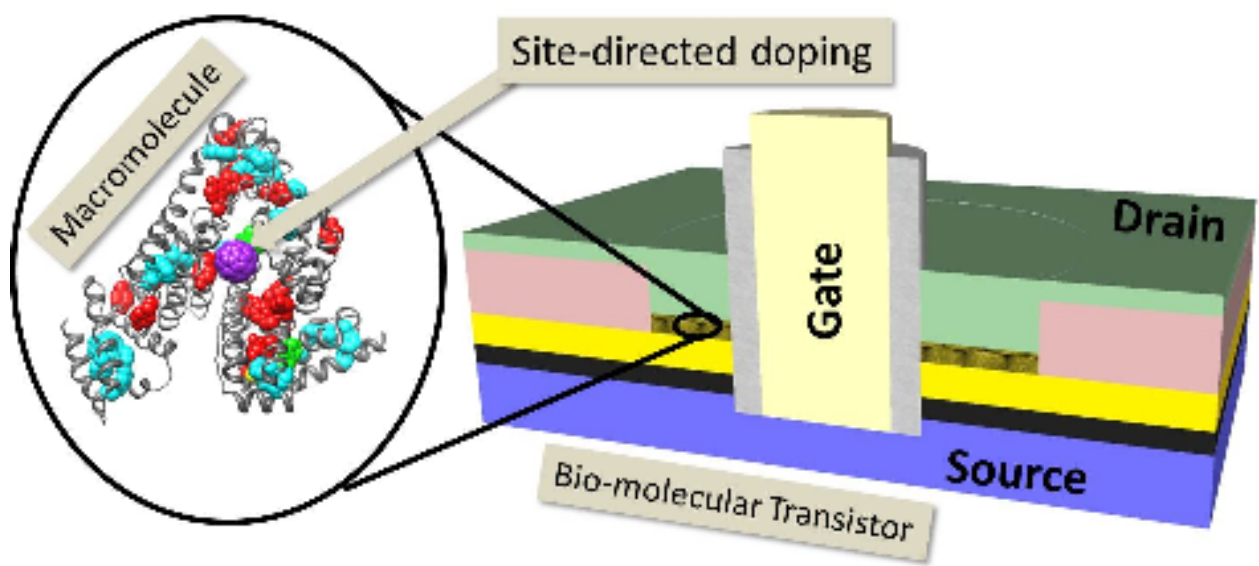